# Stacking effect and Coulomb correlation in layered charge density wave phase of 1$T$-NbS$_2$

Wei Wang,[1,2] Chen Si,[2] Wen Lei,[1,3] Feng Xiao,[1] Yunhui Liu,[1] Carmine Autieri,[4] and Xing Ming[1,5*]

1. College of Science, Guilin University of Technology, Guilin 541004, PR China

2. School of Materials Science and Engineering, Beihang University, Beijing 100191, PR China

3. Key Laboratory of Artificial Micro- and Nano-Structures of Ministry of Education and School of Physics and Technology, Wuhan University, Wuhan 430072, PR China

4. International Research Centre MagTop, Institute of Physics, Polish Academy of Sciences, Aleja Lotników 32/46, PL-02668 Warsaw, Poland

5. MOE Key Laboratory of New Processing Technology for Nonferrous Metal and Materials, Guangxi Key Laboratory of Optic and Electronic Materials and Devices, Guilin University of Technology, Guilin 541004, PR China

## ABSTRACT

Two-dimensional (2D) layered materials have attracted tremendous interest from the perspective of basic physics and technological applications in the last decade. Especially, the artificially assembled van der Waals (vdW) heterostructures and twisted 2D materials bring out fascinating properties, and render promising applications possible by engineering the stacking order. Here, based on first-principles calculations, we explored the interplay between stacking effect and electron-electron correlation in the layered vdW material of bulk 1$T$-NbS$_2$ with a 2D charge density wave (CDW) order. Without considering the Coulomb correlation, two energetically favorable out-of-plane stacking configurations are identified: one is a metallic phase with a single-layer stacking pattern, another is a band insulator with a paired-bilayer stacking configuration. Even though the Coulomb correlation is taken into account, the two energetic favorable stacking orders are still far more stable than other stacking orders. Furthermore, increasing the Coulomb interaction, the paired-bilayer stacking configuration transforms from nonmagnetic band insulator to antiferromagnetic insulator, while the single-layer stacking undergoes a Slater-Mott metal-insulator transition, which indicates the non-negligible role of electron-electron correlation interactions. In addition, the electronic structure and magnetic ground state change drastically among different stacking configurations, providing a platform to tune the electronic structures and interlayer magnetic interactions by altering the stacking order. In contrast to the widely accepted scenario of Mott localization as the driving force behind the gap formation in the CDW phase of layered transition metal dichalcogenides, our results not only highlight the crucial role of stacking order in the electronic structures of 1$T$-NbS$_2$, but also shed fresh light on the distinct effects of Coulomb interaction in different stacking arrangements.

[*] Email: mingxing@glut.edu.cn (Xing Ming)

# I. INTRODUCTION

Two-dimensional (2D) layered materials exhibit diverse properties owing to the delicate interaction between the layers held together by weak van der Waals (vdW) forces, which attract considerable attention as potential materials for promising applications.[1-3] Furthermore, abundant physical properties of the vdW layered materials can be modulated by the interlayer interactions through intercalation, pressure, strain, twist or stacking order. For example, the rotational degrees of freedom between layers can be controlled over to create new phenomena in known materials. Manipulating the twist angle between two 2D monolayers in vdW bilayer heterostructures can tune the band structures and electrical properties to realize twistronics devices.[4,5] In particular, the discovery of unconventional superconductivity, correlated insulating behavior, ferromagnetism and topological phase in the magic-angle twisted bilayer graphene have sparked researchers' unprecedented enthusiasm.[6-11]

On the other hand, lateral translation or sliding of one layer relative to another layer can modify the stacking order and stacking symmetry of the vdW layered materials, which gives rise to rich phase diagrams and novel properties, such as the stacking-engineered ferroelectricity in bilayer BN [12,13], stacking-tunable magnetism in bilayer $CrI_3$ and $CrBr_3$ [14,15], and stacking-driven metal-insulator transitions in layered $1T$-$TaS_2$ [16-19]. $1T$-$TaS_2$ is an archetypal charge density wave (CDW) material that has been in the spotlight for many years, which transforms into a commensurate CDW (CCDW) phase characterized by $\sqrt{13} \times \sqrt{13}$ star-of-David (SD) clusters (schematically shown in **Figure 1**) upon cooling to ~200 K. Recently, rekindled interest in $1T$-$TaS_2$ stems from the extensive debate on the origin of the insulating nature in the CCDW phase. According to traditional wisdom, there is one unpaired electron forming a half-filled band in the CCDW phase, therefore, the insulating ground state is driven by the Mott-Hubbard mechanism.[20] However, first-principles density functional theory (DFT) calculations combined with transport measurements, angle-resolved photoemission spectroscopy (ARPES), low-energy electron diffraction experiments and atomic resolution scanning transmission electron microscopy (STEM) give robust evidence that the insulating gap of the CCDW phase is primarily governed by the out-of-plane stacking arrangements of the SD clusters rather than the electron-electron correlation.[16-19,21-29] These theoretical and experimental results highlight the critical role played by the stacking order of the SD clusters intertwining with orbital order in determining the electronic structure of the CCDW phase.

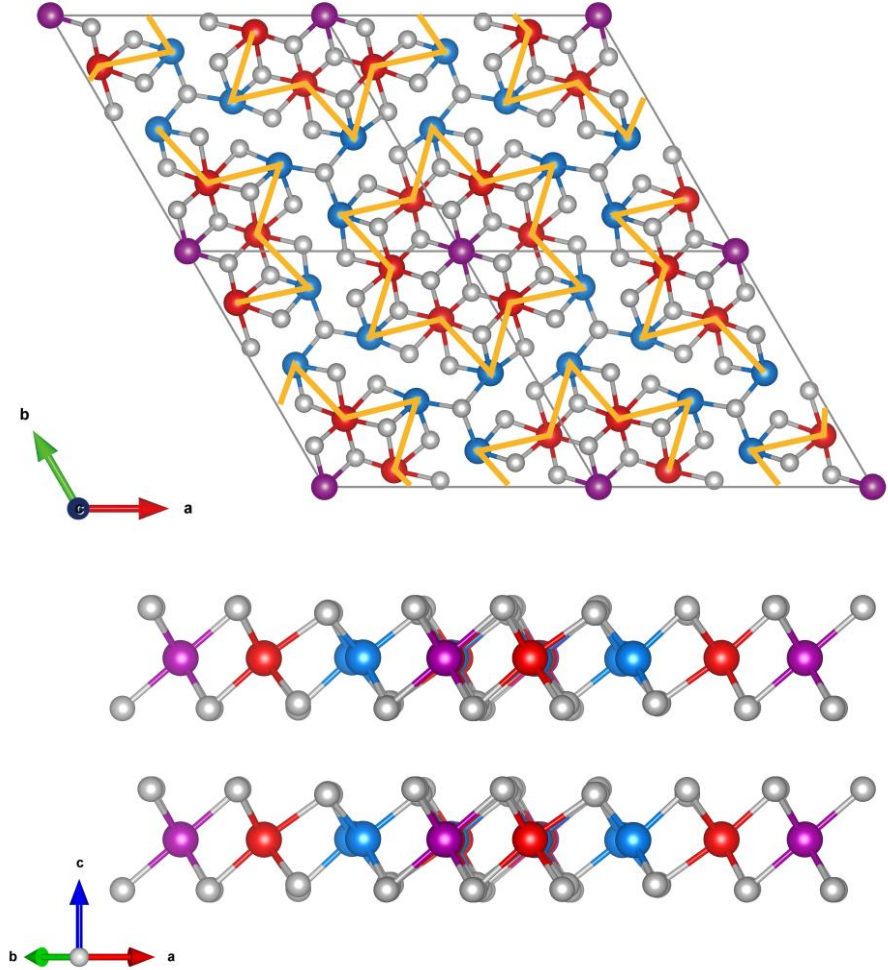

**Figure 1** Top and side views of the CCDW phase of $1T$ polytype $TaS_2$ or $NbS_2$ characterized by the $\sqrt{13}\times\sqrt{13}$ SD clusters (schematically drawn in yellow line). The big balls denote the Ta or Nb atoms (the purple, red and blue colors indicate three different structural symmetric sites), and the small balls (gray color) represent the S atoms.

Recently, we and Tresca *et al*. theoretically predicted that $1T$-$NbS_2$ belonging to CDW materials, which is isostructural and isovalent to $1T$-$TaS_2$ and also exhibits the typical $\sqrt{13}\times\sqrt{13}$ SD structural reconstructions (**Figure 1**).[30,31] In the bulk phase, the unit cell of $1T$ polytype $NbS_2$ consists of corner-sharing monolayer $NbS_6$ octahedra, which differs from the bilayer and trilayer stacking arrangements of the corner-sharing trigonal prismatic coordination $NbS_6$ layers in $2H$ and $3R$ polytypes $NbS_2$.[32-34] The distinct stacking arrangements of the coordination polyhedra lead to dramatically different electronic structures ranging from insulator to metal to superconductor for these three polytypes of layered $NbS_2$.[30-36] The monolayer $1T$-$NbS_2$ exhibits a ferromagnetic (FM) spin ½ insulating state,[31] while an interlayer antiferromagnetic (AFM) ordering may dominate the interlayer coupling in the bulk phase [30]. Diffuse X-ray scattering measurements detect traces of CDW in 2$H$-NbS$_2$, which is also attributed to the local 1$T$ polytype environment at the interface of the rotational stacking faults between 2$H$ domains.[34] Martino *et al.* propose that the magnetic CDW state of the 1$T$-NbS$_2$ layers can intercalate the host of 2$H$ stacking NbS$_2$ and form a 2$H$/1$T$-NbS$_2$ heterostructure, and result in an unexpected unidirectional Kondo scattering in metallic 2$H$-NbS$_2$.[36] The stacking order is expected to create lateral shifts of the CDW clusters between adjacent layers and provide an effective route to manipulate the electronic phases of the layered 1$T$-NbS$_2$. Furthermore, being the same as the CCDW phase of 1$T$-TaS$_2$, the unit cell of 1$T$-NbS$_2$ is reconstructed into a triangular lattice with prototypical $\sqrt{13} \times \sqrt{13}$ SD clusters, and there is a single unpaired electron with $S = ½$ spin moment localizing in the centers of each SD cluster.[30,31] Therefore, exotic quantum phases, such as quantum spin liquid, are expected to emerge in these triangular-lattice materials.[37-42]

The motivation of the present paper is to further examine the out-of-plane stacking order of the SD clusters in the CCDW phase by first-principles DFT calculations, and uncover the role played by the stacking order and electron-electron correlation in the stability and electronic phase of 1$T$-NbS$_2$. We not only unveil the tunable electronic structures derived from the distinct stacking order of the SD clusters, but also reveal a significant effect of the Coulomb correlation in manipulating the electronic properties of the CCDW phase 1$T$-NbS$_2$. We hope our theoretical results will stimulate further experimental and theoretical studies on the electronic properties of the CCDW phase to better understand the intricate interplay between the stacking order and electron correlation in 1$T$-NbS$_2$. Furthermore, although we examine the case of the 1$T$-NbS$_2$, similar effects could be present in other members of the same material class.

## II. STRUCTURE MODELS AND COMPUTATIONAL DETAILS

The routinely stacked CCDW phase of 1$T$-NbS$_2$ can be described by the space group of P$\bar{3}$. Each Nb-atom is bonded to six S-atoms forming distorted octahedral coordination, and the adjacent layers are bonded together by weak vdW forces (the schematic model is presented in **Figure 1**). The building block of the CCDW phase is the SD cluster as shown in **Figure 2(a)**. Due to the threefold in-plane rotational symmetry of the CCDW phase, five kinds of Nb-atoms can be defined, where the central Nb-atom is labeled as 0 and the surrounding Nb-atoms are labeled as 1 to 4 in each SD cluster, respectively. Taking the threefold in-plane symmetry into account, the out-of-plane alignments of the SD clusters in adjacent planes are described by five types of stacking interfaces $t_i$ (i = 0, 1, 2, 3, 4) [illustrated in **Figures 2(b)-(f)**], which lead to five types of out-of-plane 3D stacking order $T_i$ (i = 0, 1, 2, 3, 4). The routinely stacking configuration $T_0$ corresponds to the on-top stacked monolayer with the stacking interface $t_0$ [**Figures 2(b)**], whereas other single-layer stacking configuration $T_i$ can be viewed as a lateral shift [the dashed line depicted in **Figures 2(c)-(f)**] of every single layer with respect to the adjacent layer with the corresponding stacking interface $t_i$. Accordingly, the central Nb atoms of the SD in the upper layer are located above the Nb atom in one of the tips ($t_1$ and $t_4$ interfaces) or the inner circle hexagon ($t_2$ and $t_3$ interfaces) of the SD of the lower layer. In addition, we also considered paired-bilayer stacking configurations with a lateral sliding of every two layers due to the possibility of the Peierls dimerization for a chain of half-filled bands. These four types of stacking order $T_{0i}$ (i = 1, 2, 3, 4) can be viewed as on-top stacked bilayers with the stacking interface $t_0$, which by themselves are stacked by a stacking interface $t_i$ (i = 1, 2, 3, 4) in the out-of-plane direction.[16,17,21,26] All these stacking orders are illustrated detailly in **Figure 3**.

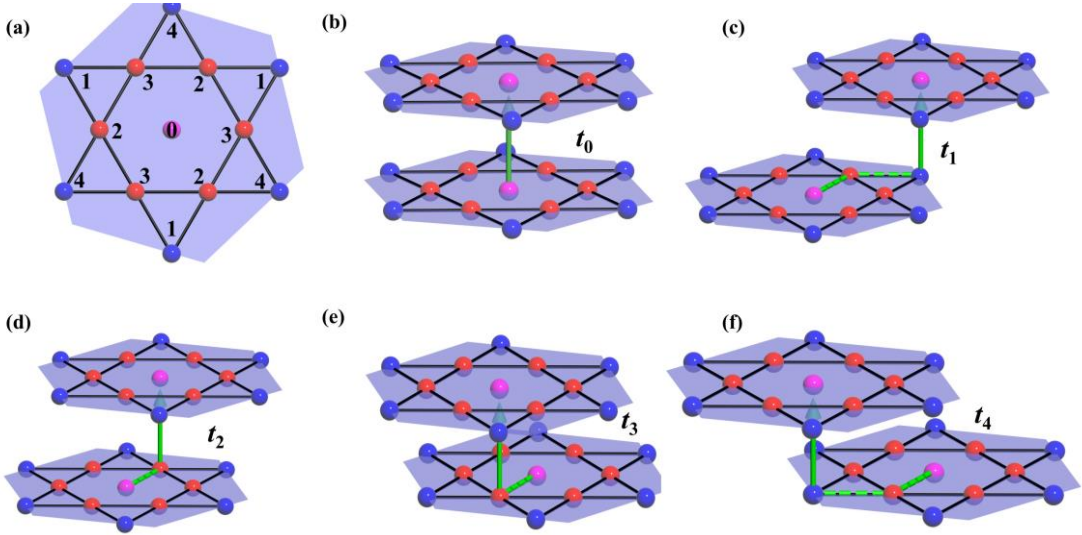

**Figure 2** Schematic illustration of the $\sqrt{13} \times \sqrt{13}$ SD cluster in the CCDW phase of $1T$ polytype NbS$_2$ and different out-of-plane stacking order (the S atoms are not shown). (a) In-plane SD cluster and classification of the Nb-atoms. (b)-(f) are five kinds of stacking interfaces ($t_0$-$t_4$) between two adjacent layers. Solid green arrows illustrate the Nb-atom in the first layer facing to the central Nb-atom in the adjacent layer, and dashed green lines denote the in-plane lateral shift of the SD clusters with respect to the counterparts in the adjacent layer.

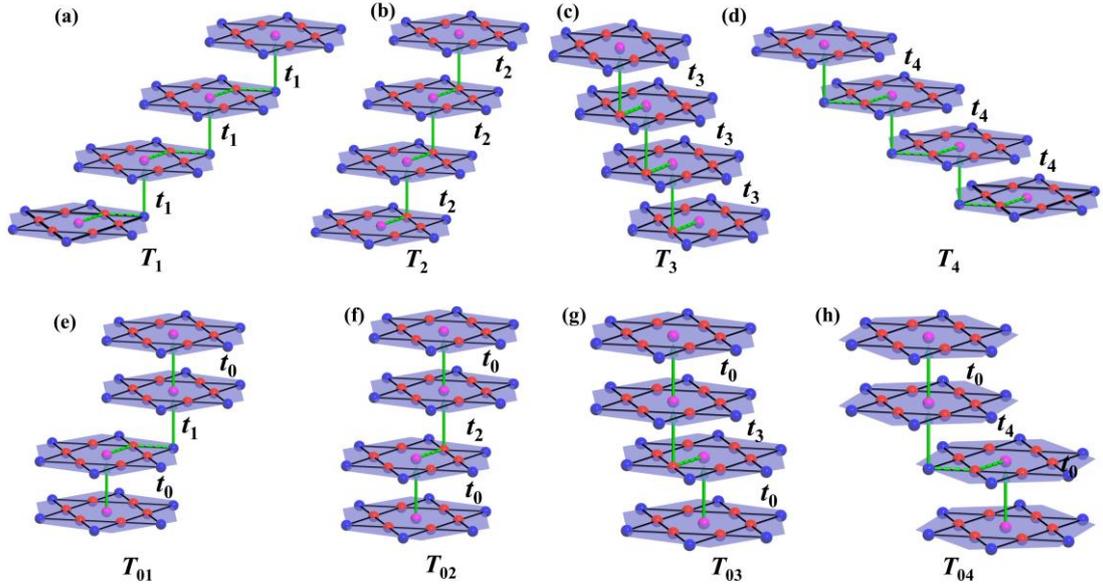

**Figure 3** Schematics of the out-of-plane stacking order (the S atoms are not shown) of the SD clusters in the CCDW phase of 1$T$-NbS$_2$: (a)-(d) are four kinds of stacking patterns of $T_1$-$T_4$ with corresponding interlayer interface $t_i$ ($t_1$-$t_4$) between two adjacent layers, (e)-(h) are four kinds of bilayer stacking patterns of $T_{01}$-$T_{04}$ with corresponding interlayer interface $t_i$ ($t_1$-$t_4$) between two adjacent on-top stacked bilayers with $t_0$ interface.

Based on DFT, we performed all of the calculations by using Vienna *ab initio* simulation package (VASP) [43,44] within the generalized gradient approximation (GGA) [45] approach and projector-augmented wave (PAW) [46] potentials. The kinetic-energy cutoff of 520 eV was used, and Brillouin zone (BZ) was segmented by Gamma centered grid of 0.2 Å$^{-1}$ for all structures.[47] Considering the weak interlayer interactions in the CCDW phase, vdW dispersion correction of DFT-D3 [48] was employed to perform geometrical optimization so that accurate lattice information and total energy can be obtained. Among the members of the $MX_2$ ($M$ = Nb, Ta; $X$ = S, Se) family, the NbS$_2$ is composed of the lighter atoms and therefore is the most correlated one. To deal with the electronic correlation, we add the Coulomb repulsion $U$ on the 4$d$ orbitals of the Nb ranging from 0 to 3.5 eV that are typical values for 4$d$ orbitals.

### III. RESULTS AND DISCUSSIONS

Firstly, the normal on-top vertically stacking order $T_0$ of the CCDW phase [**Figure 2(b)**] serves as the starting point of our study. The optimized lattice parameters are $a$ = 12.129 Å and $c$ = 5.926 Å at the DFT-D3 level, which are consistent with previous results.[30] In agreement with our previous results,[30] without electronic correlations the CCDW phase always converges to a nonmagnetic (NM) metallic state independent on the preset magnetic configuration. When we add the Coulomb interaction, the magnetic ground state is an interlayer AFM state that simultaneously opens the band gap [**Figure S1** in the Supplemental Material (SM)].[49] However, as shown in **Table 1**, the calculated results indicate that the stacking order plays an essential impact on the layer spacing and stability. The total energy of the system depends remarkably on the stacking order and varies up to ~60 meV/SD. Compared with the $T_0$ stacking order, the layer spacing shrinks a bit and results in significant changes of the vdW energy, where the vdW energy $\Delta E_{vdw}$ is positively proportional to the change of the layer spacing $\Delta c$. The vdW energy decreases as the layer spacing shrinks, demonstrating the crucial role of the vdW interactions is to anchor the layers.[50]

**Table 1** Equilibrium layer spacing ($c$) and energy landscape of the different stacking configurations for the NM states calculated within GGA. The changes of the layer spacing ($\Delta c$), total energy ($\Delta E_{tot}$), and vdW energy ($\Delta E_{vdw}$) are calculated with respect to the $T_0$ stacking.

| Stacking order | $c$ (Å) | $\Delta c$ (%) | $\Delta E_{tot}$ (meV/SD) | $\Delta E_{vdw}$ (meV/SD) |
|---|---|---|---|---|
| $T_0$ | 5.926 | 0 | 0 | 0 |
| $T_1$ | 5.887 | -0.658 | -24.82 | -75.77 |
| $T_2$ | 5.910 | -0.270 | 6.45 | -29.72 |
| $T_3$ | 5.909 | -0.287 | 12.11 | -31.28 |
| $T_4$ | 5.923 | -0.439 | 27.59 | -46.79 |
| $T_{01}$ | 5.908 | -0.304 | -36.29 | -36.54 |
| $T_{02}$ | 5.920 | -0.101 | -4.57 | -11.58 |
| $T_{03}$ | 5.920 | -0.101 | 0.21 | -10.96 |
| $T_{04}$ | 5.923 | -0.051 | -5.51 | -4.87 |

Furthermore, two stacking configurations of $T_1$ and $T_{01}$ exhibit remarkable stability from an energetic viewpoint. $T_1$ stacking order is the most favorable stacking arrangement for the vdW energy. On the contrary, although the benefit of vdW energy is not significant, $T_{01}$ stacking is preferred by total energy. Both $T_1$ and $T_{01}$ stacking orders are featured by the highly stable $t_1$ interface over the other interfaces, which are closely related to the favorable van der Waals interaction and interlayer S–S bonding.[21] In fact, the interlayer S-atoms configurations in different stacking systems are distinct from each other upon the specific stacking order. Shorter or longer S-S distances can be observed in the layer spacing. Such a steric effect of the S-atoms plays a significant impact on the energy of the stacking system. A shorter S–S distance results in an interlayer bonding with the $3p_z$ orbitals, which is favorable for energy, but such a bonding interaction is disallowed for a longer S-S distance. As exemplified by $T_1$ and $T_4$ stacking orders, we can clearly identify the different interlayer distances between the nearest neighbor S-atoms coordinated with the central Nb-atoms of the SD cluster [**Figures 4(a)** and **(b)**]. As shown in the corresponding charge density maps [**Figures 4(c)** and **(d)**], bonding behavior can be clearly observed for the $T_1$ stacking with an interlayer S-S distance of ∼3.29 Å, whereas no such trace for the $T_4$ stacking with the interlayer S-S distance of ∼4.71 Å.

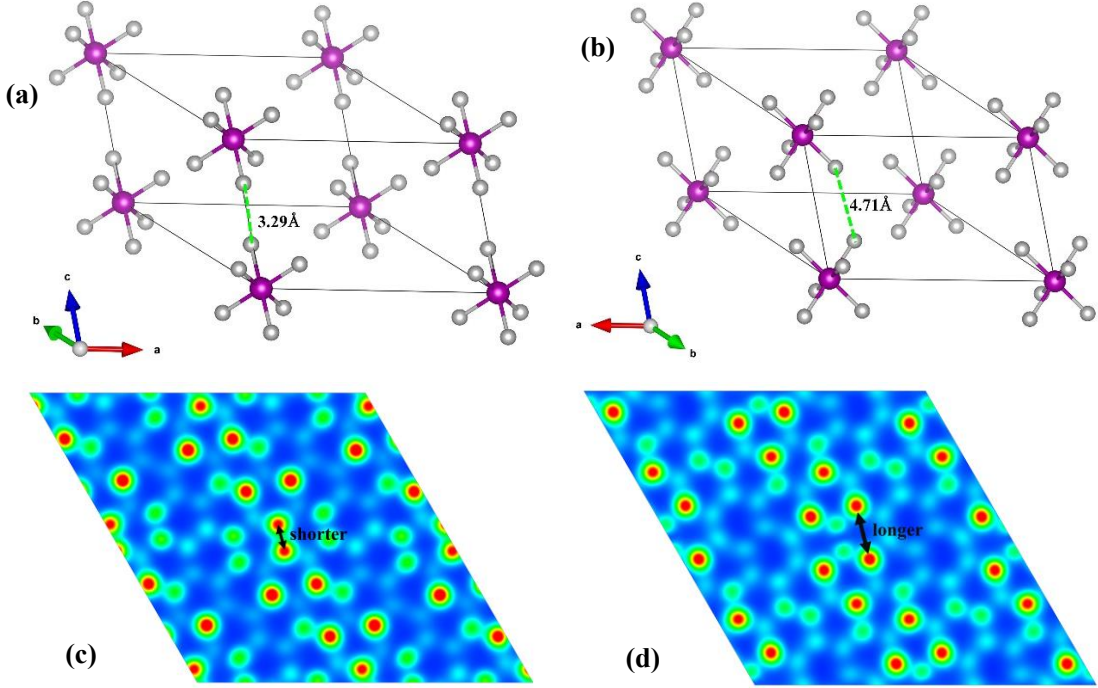

**Figure 4** (a) and (b) are side views of the $T_1$ and $T_4$ stacking order (only the central Nb-atoms of the SD cluster and the coordinated S-atoms are shown). (c) and (d) are charge density maps corresponding to the $T_1$ and $T_4$ stacking order, respectively, obtained by integrating electronic states near the Fermi level. The slices are located at the center of $t_1$ and $t_4$ interfaces and parallel to the *ab* plane.

In addition, the electronic structures show obvious dependence on the stacking arrangements of the CCDW phase (see **Figure 5** and **Figure S2** in the SM).[49] In particular, the relative position of the SD clusters in adjacent *ab*-planes has a significant influence on the band dispersions. All the $T_i$ stacking patterns show metallic behavior, whereas the $T_{0i}$ stacking arrangements show striking semiconductor–metal transitions. As shown in **Figure 5(a)**, the $T_1$ stacking is metallic with a half-filled band characterized by the $4d_{3z^2-r^2}$ orbital contribution from the central Nb-atom of the SD clusters. Due to the presence of the $4d_{3z^2-r^2}$ bands at the Fermi level, the electronic properties strongly depend on the stacking along the *z*-axis. Attributed to the interlayer S-S bonding interactions [**Figure 4** (a)], the $T_1$ stacking order shows a significant in-plane hopping [17] and stronger three-dimensional metallicity, where the bands cross the Fermi level not only along Γ–*A* but also along *L*–*A* and *L*–*H*. By contrast, the special orbital order intertwined with the CDW in the $T_0$ stacking only permits significant out-of-plane charge hopping, corresponding to the in-plane insulating characteristics along Γ–*M*–*K* and out-of-plane one-dimensional metallic band dispersion along Γ–*A*.[30] Furthermore, as shown in **Figure 5(b)**, an inherent band gap of 45.9 meV is opened up for the $T_{01}$ stacking within GGA approach, implying a band insulating behavior instead of Mott insulator. Each cell is constructed from two SD clusters, and a pair of degenerate bands is almost completely isolated in the uppermost valence band. The contributions of the $4d_{3z^2-r^2}$ characteristics from the two central Nb-$0_1$ and Nb-$0_2$ atoms of the two SD are almost the same, indicating a degeneracy of them. A cell with an even number of electrons can yield filled valence bands forming an insulator or semiconductor.[51] Obviously, two orphan electrons from Nb-$0_1$ and Nb-$0_2$ pair with each other and fill the isolated orbit, accompanied by an opening of the band gap and a lowering of the total energy. Therefore, the stacking arrangements enable one to manipulate the band dispersion and gap structure of the CCDW phase 1*T*-NbS$_2$. Similar to 1*T*-TaS$_2$, the stacking order yields a new device concept, which utilizes metastable stacking orders to control the electronic structure of nanostructures and causes a semiconductor–metal transition.[17]

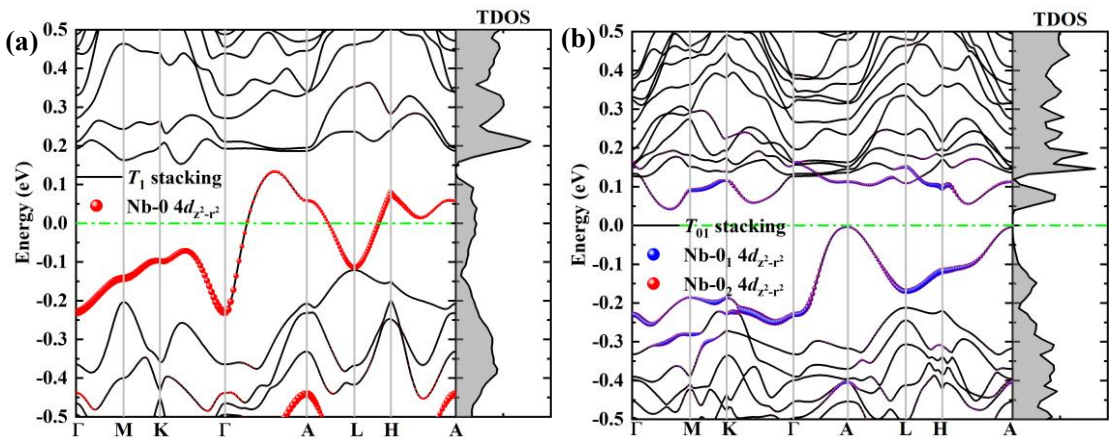

**Figure 5** Band structures and corresponding density of states of the (a) $T_1$ and (b) $T_{01}$ stacking within GGA approach. The red balls in (a) denote the orbital contributions with $4d_{3z^2-r^2}$ character of the central Nb atom of the SD cluster (Nb-0). The blue/red balls in (b) denote the orbital contributions with $4d_{3z^2-r^2}$ character of Nb-$0_1$/Nb-$0_2$ (central Nb atom of the SD in the first/second layer).

Attributed to the stacking effects, a band insulator is identified by interlayer dimerization in $T_{01}$ stacking of the CCDW phase 1$T$-NbS$_2$, which is in contrast to the Mott-Hubbard mechanism proposed in previous studies on the origin of the insulating nature in the CCDW phase of 1$T$-TaS$_2$ and monolayer 1$T$-NbS$_2$.[23,31] However, recent studies on 1$T$-TaS$_2$ have also implied that this dimerization cannot be simply depicted by a genuine Peierls mechanism if the continuity of order parameter has been taken into account, which is strongly related to the interlayer hopping.[16-18,21,22,24] It is worth noting that the $T_{01}$ stacking consists of two interfaces: the $t_0$ interface inside each on-top stacked paired bilayer and the $t_1$ interface between the adjacent bilayers. Once a stacking order is established, the interlayer hopping integral between adjacent layers will be changed dramatically, which has an essential impact on the electronic structure. The interlayer hopping in the $t_0$ interface is the same in the $T_0$ stacking, while interlayer hopping is much greater in the $t_0$ interface than the $t_1$ interface in the $T_{01}$ stacking.[18] Moreover, there exists a continuous crossover region between the band insulating and Mott insulating phase in the bilayer Hubbard system, and the stronger interlayer hopping is beneficial to the band insulator.[52,53] Similar to the previous findings for 1$T$-TaS$_2$, the strong electron-electron correlation does not play a major role in the insulating nature of the $T_{01}$ stacking, whereas the on-top stacked bilayer and the interlayer hopping should be the most important factors.[16,21]

Despite the band insulating behavior can be realized by paired SD layers without considering the electron-electron correlation,[16,17,21] the strong correlations and Mott physics still exist in the limit of monolayers.[31] Furthermore, the orphan electron on each SD cluster provides $S = ½$ spin moment in the triangle lattice, which offers an opportunity to study the Coulomb correlation and quantum state in the CCDW phase of bulk 1$T$-NbS$_2$. By considering Coulomb correlations at the level of $U = 2.95$ eV,[30] we calculate the energy differences between different stacking configurations to explore the effects of the electron-electron correlation on the relative stability and the electronic structures. Especially, by doubling the cell along the layer stacking direction, two magnetic ordering states of AFM↑↓ and FM↑↑ are preset for $T_i$ stacking, whereas four magnetic states of AFM↑↓↑↓, AFM↑↓↓↑, AFM↑↑↓↓ and FM↑↑↑↑ are considered for $T_{0i}$ stacking. As shown in **Figure 6**, the magnetic ordering states exhibit lower energy relative to the NM state for all stacking patterns, which implies possible magnetic states at low temperature. The unpaired single-layer $T_i$ stacking patterns, especially the $T_1$ and $T_4$ stacking arrangements show a stronger response to the magnetic ordering and Coulomb correlation. By contrast, the energy gains of the $T_{0i}$ stacking are not so striking, that the energy of the AFM state and the NM state differs only a few meV/SD. Moreover, the distinct stacking arrangements and interfaces show different magnetic configurations. For instance, the $T_4$ stacking tends to form FM↑↑ state rather than AFM↑↓ state, and $T_{04}$ stacking tends to form AFM↑↓↓↑ state rather than AFM↑↓↑↓ state, implying that the $t_4$ interface is prone to form FM↑↑ configuration. In accord with the $T_0$ stacking, the $t_0$ interface tends to form AFM paired bilayers, all the $T_{0i}$ stacking orders are featured with AFM↑↓ configuration in the $t_0$ interface. Therefore, a variation of the stacking orders by a lateral sliding could change the direct hopping strength between interlayer S orbitals, which further alters the interlayer magnetic interactions between FM and AFM ones. The stacking order provides an effective way to tailor the interlayer magnetic interactions, which opens a paradigm for seeking metamagnetic 2D materials from AFM to FM state in real applications.[14,15]

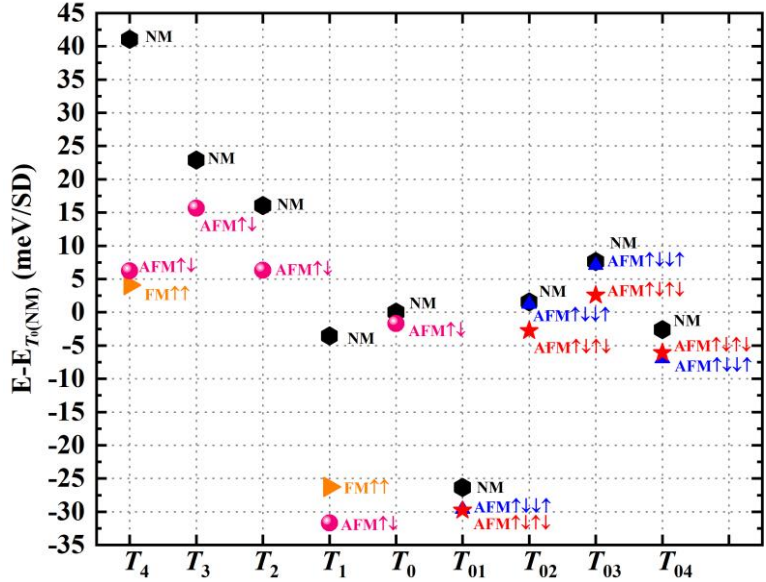

**Figure 6** Energy landscape of different stacking order with respect to $T_0$ stacking of the NM state. All magnetic ordering states are calculated by GGA + $U$ within the Coulomb correlation of $U$ = 2.95 eV. The upward and downward arrows represent the directions of the spin magnetic moments in different layers. We should note that not all our preset magnetic configurations are presented here, because the unstable magnetic states have converged to NM state or other states.

The $T_1$ and $T_{01}$ stacking configurations are still far more stable than the other stacking orders (by ~30 meV/SD) (**Figure 6**). Surprisingly, the $T_1$ stacking obtains a huge energy gain after entering the magnetic ordering states, so that the total energy is lower than that of the $T_{01}$ stacking by about 1.9 meV/SD. In addition, in contrast to the metallic band structure of the NM state, a tiny insulating gap of 12.9 meV is opened up for the AFM state (**Figure 7**), which signifies the Mott-Slater insulating nature of the $T_1$ stacking (will be discussed in detail later). The electron-electron correlation and magnetic interactions play a minor impact on the $T_{01}$ stacking, the energy differences between the NM and AFM states are within only 3 meV/SD. The two AFM states (AFM↑↓↓↑ and AFM↑↓↑↓) of the $T_{01}$ stacking order almost have identical energies, indicating that the $T_{01}$ stacking is not sensitive to the interlayer magnetic configurations, and the typical characteristics of the band insulator are robust. This phenomenon is consistent with the recent experimental confirmation that the paired bilayer $T_{01}$ stacking is the dominant building block of $1T$-TaS$_2$ at low temperatures.[18,24] As stated above, the primary origin of the insulating properties in $T_{01}$ stacking is the dimerization and interlayer hopping, while strong electron-electron correlation does not play a key role. In other words, $T_{01}$ stacking is insensitive to Coulomb repulsion $U$, in contrast to the case of the $T_1$ stacking.

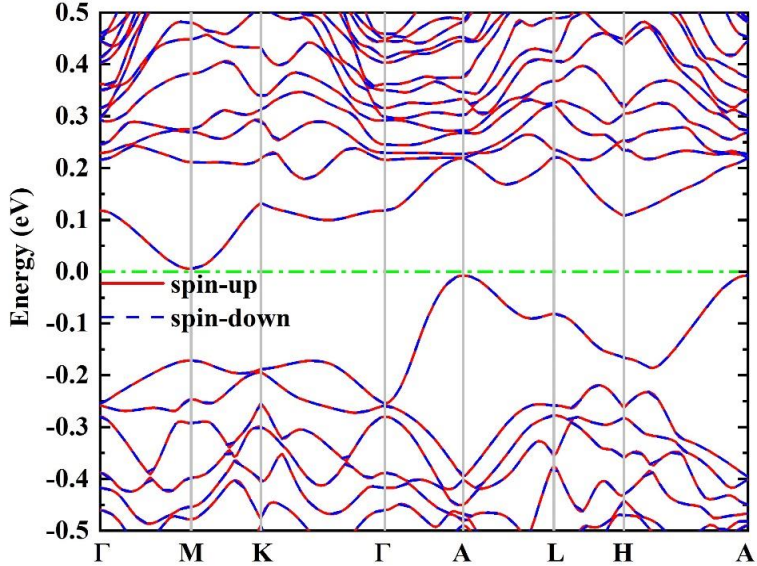

**Figure 7** Band structure of the interlayer AFM state for $T_1$ stacking calculated by GGA + $U$ ($U$ = 2.95 eV).

Compared with the $T_{01}$ stacking order, the distinct insulating mechanisms leads to the compensation of the total energy for $T_1$ stacking under the action of $U$, which eventually results in a lower-energy stacking order. Notable is that the total energy difference between the $T_1$ and $T_{01}$ stacking configurations is only ~2 meV/SD. The CCDW phase of bulk $1T$-NbS$_2$ provides a platform to reveal the underlying role of $U$ in the ground state with different stacking orders.

Considering that the Nb element has a 4$d$ shell with moderate electron-electron correlation, we choose the $U$ value ranging from 0 eV to 3.5 eV. As shown in **Figure 8**, the $T_{01}$ stacking order converges to NM insulating states when the $U$ values are lower than 2.5 eV, then the NM insulating state transforms to an AFM insulating state when the $U$ value reaches up to 2.5 eV. Dynamical mean-field theory studies on the bilayer Hubbard model have identified a crossover from band insulator to Mott insulator under the effect of increasing $U$ values.[52,53] On the contrary, the $T_1$ stacking order shows lower-energy NM metallic states when the $U$ values are lower than 1.5 eV, and the energy differences between the $T_1$ and $T_{01}$ stacking orders gradually increase with the increasing $U$ value. Once the $T_1$ stacking order transforms from the NM metallic to AFM metallic states at $U = 1.75$ eV, the evolution trend of the energy differences shows an upturn, which gradually decreases with the increasing $U$ values. When the $U$ value reaches up to 2.5 eV, the $T_1$ stacking order transforms from the AFM metallic state to an AFM insulating state. Therefore, both Coulomb repulsion and magnetic order are necessary to open the gap in the $T_1$ stacking order. As shown in other compounds, this is a signature of a Slater-Mott insulator.[54-56] We can claim that along with the increasing $U$ values, the $T_1$ stacking order exhibits an interesting Slater-Mott metal-insulator transition from NM metal to AFM metal and finally to AFM insulator. Along with the $U$ value further increasing, the energy differences between the $T_1$ and $T_{01}$ stacking orders gradually decrease. As a result, the ground state transforms from the band insulator to the Slater-Mott insulator. Previous calculations suggest that the band insulator may be so fragile that electronic doping or applying strain can easily destroy the relative stability of the two stacking configurations.[21] We speculate that the CCDW phase of bulk 1$T$-NbS$_2$ locating at the crossover region of Slater-Mott insulator and band insulator, remaining to be uncovered experimentally. More precisely, both Coulomb repulsion and magnetism are needed to open the bandgap making these states a Slater-Mott insulator.

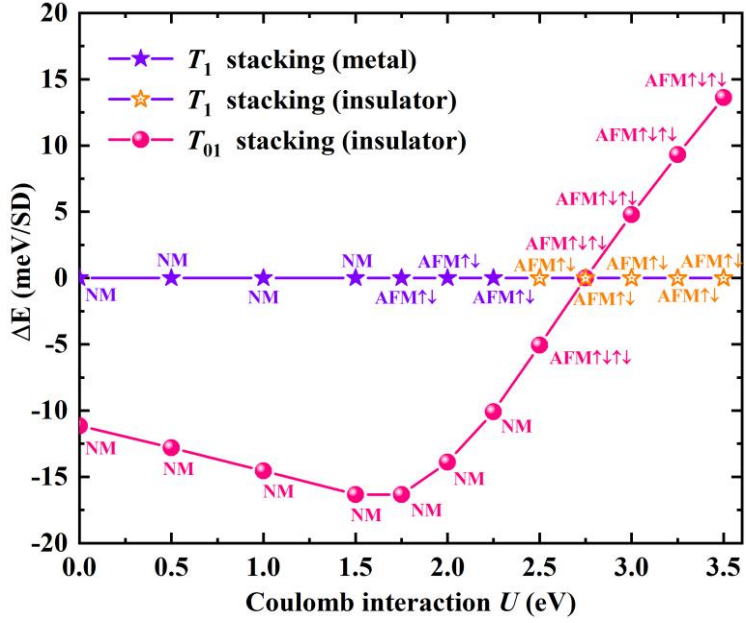

**Figure 8** The evolutions of the energy differences and metal-insulator transitions between $T_1$ and $T_{01}$ stacking along with the increasing Coulomb correlation parameters $U$.

The bilayer-stacked bulk phase of 1$T$-NbS$_2$ and 1$T$-TaS$_2$ may satisfy the criteria of a simple band insulator, but ARPES and X-ray diffraction data have suggested certain randomness of the out-of-plane stacking order, and STEM topographic images show the broken dimerization by unpaired SD clusters at the surfaces of the CCDW phase 1$T$-TaS$_2$.[18,24] These experiments demonstrate the possible existence of an unpaired stacking SD cluster accompanied with a small change of stacking order, which will yield band insulator to Mott insulator transitions and metal-insulator transitions. The small energy differences between the $T_1$ and $T_{01}$ stacking order jointed with the possible stacking faults and stacking disorder in real materials will give rise to complex phase diagrams for the CCDW phase of 1$T$-NbS$_2$. Additionally, the electron-electron correlation cannot be absolutely precluded, Mott physics might still play an important role in the electronic structure of the CCDW phase.[16] We note that first-principles calculations within traditional DFT, DFT + $U$ and the hybrid functional have failed to reproduce the out-of-plane insulating behavior of the bulk 1$T$-TaS$_2$ for the $T_0$ stacking order with on-top stacked SD clusters.[16,17,21,25,57] Recently, a Mott insulating band structure with the interlayer AFM order has been successfully achieved by applying the $U$ potential onto a generalized basis.[58] However, any magnetic ordering has never been observed in 1$T$-TaS$_2$ so far.[39-42] In contrast, the calculated electronic structures show a band insulating behavior without considering the magnetic configurations in the CCDW phase of 1$T$-NbS$_2$, which transforms to Slater-Mott insulating state by considering the magnetic ordering states and Coulomb correction with DFT + $U$. Furthermore, the energy differences of the magnetic ordering states between the most stable $T_1$ and $T_{01}$ stacking order are very small, which implies a possible magnetic disordering associated with the triangular lattice. These stacking-driven metal-insulator transitions are similar to thickness-driven metal-insulator transitions in transition-metal oxides, where in both cases the structural effects are responsible for a bandwidth reduction and consequently for a thickness-controlled metal-insulator transition towards more correlated electronic phases.[54-56] The presence of different electronic and magnetic phases paves the way to manipulate the competition between these phases by doping or applied external pressure. The concomitant insulating nature, localized orphan spin, and possible magnetic disordering on the triangular lattice of CCDW phase 1$T$-NbS$_2$ will induce exotic quantum states, perhaps the elusive quantum spin liquid, needing to be further clarified both theoretically and experimentally.

## IV. CONCLUSIONS

In summary, we explored the influences of stacking order and electron correlation on the electronic properties of the CCDW phase 1$T$-NbS$_2$ by first-principles calculations. The electronic structures show a significant dependence on the stacking order. Without considering the electron-electron correlation, two energetic favorable stacking orders of $T_1$ and $T_{01}$ show distinct electronic structures, one is metal and the other one is band insulator. These two energetic favorable stacking orders are still robust while considering the electron-electron correlation. Particularly, the ground state of the system is strongly related to the strength of the electron-electron correlation. Along with the increasing Coulomb interactions, the $T_1$ stacking order undergoes a Slater-Mott metal-insulator transition, whereas the $T_{01}$ stacking order transforms from band insulator to AFM insulator, indicating the indispensability of electron correlation. Our work highlights the critical role of interlayer stacking order and reveals an important role of the Coulomb correlation in the electronic properties of the CCDW phase 1$T$-NbS$_2$. Furthermore, we propose stacking-controllable metal-insulator transitions, metamagnetic transitions, and exotic quantum phase perhaps existing in the promising 1$T$-NbS$_2$, which will draw intensive attention.

## ACKNOWLEDGMENTS

We would like to thank the fruitful discussion with Dr. Sung-Hoon Lee from Kyung Hee

University. The work was sponsored by the National Natural Science Foundation of China (No. 11864008) and Guangxi Natural Science Foundation (No. 2018AD19200 and 2019GXNSFGA245006). C. A. is supported by the Foundation for Polish Science through the International Research Agendas program co-financed by the European Union within the Smart Growth Operational Programme.